\documentclass{ar-1col-S2O}
\usepackage[numbers]{natbib}
\usepackage[T1]{fontenc} 
\usepackage{url}
\jname{Xxxx. Xxx. Xxx. Xxx.}
\jvol{AA}
\jyear{YYYY}
\doi{10.1146/annurev-nucl-121423-100906}

\begin{document}

\markboth{K.~Bloom and V.~Boisvert}{Sustainability of Accelerators}

\title{Sustainability and Carbon Emissions of Future Accelerators}

\author{Kenneth Bloom$^1$ and V\'{e}ronique Boisvert$^2$
\affil{$^1$Department of Physics and Astronomy, University of Nebraska-Lincoln, Lincoln, NE, USA 68588-0299; email: kenbloom@unl.edu}
\affil{$^2$Department of Physics, Royal Holloway University of London, Egham, Surrey, TW20 0EX, UK; email: veronique.boisvert@rhul.ac.uk}}

\begin{abstract}
Future accelerators and experiments for energy-frontier particle physics will be built and operated during a period in which society must also address the climate change emergency by significantly reducing emissions of carbon dioxide.  The carbon intensity of many particle-physics activities is potentially significant, such that as a community particle physicists could create substantially more emissions than average citizens, which is currently more than budgeted to limit the increase in average global temperatures.  We estimate the carbon impacts of potential future accelerators, detectors, computing, and travel, and find that while emissions from civil construction dominate by far, some other activities make noticeable contributions.  We discuss potential mitigation strategies, and research and development activities that can be pursued to make particle physics more sustainable.  
\end{abstract}

\maketitle

\tableofcontents

\section{INTRODUCTION}
Particle physics seeks to understand the universe at the level of its fundamental constituents.  The United States particle physics community, through the Particle Physics Project Prioritization process (P5) has recently developed a new strategic plan that sets out six science drivers for the field, a series of new experimental facilities that addresses them, and a program of research and development of new technologies that would allow the exploration of extremely small distance, and thus high energy, scales~\cite{Murayama_2023}.  Among these facilities are new particle accelerators larger than any built to date, and  R\&D efforts could open the possibility of even larger machines.  The P5 plan has generated significant excitement in the community, as it sets out a compelling scientific path for the coming decades, and includes the possibility of constructing the first energy-frontier collider in the U.S. since the closure of the Fermilab Tevatron in 2011. On the European side, CERN has recently launched the next European Strategy Update~\cite{ESUPP} that will set out the priorities of the field in Europe for the next five years. 

However, over these same decades, our planet will be threatened by human-induced climate change at a scale that could disrupt our scientific plans.  Should global warming exceed 1.5$^\circ$C there will be significant threats to human well-being due to reduced food and water supplies, increased rates of disease, and decreased availability of land due to sea level rise~\cite{IPCC-sr15}.  The societal challenges arising from this so-called ``climate emergency''~\cite{ClimateEmergency} could draw resources away from fundamental research, no matter how compelling the scientific case.  The sixth assessment report of the Intergovernmental Panel on Climate Change has concluded that it is ``unequivocal'' that global warming is due to the emission of carbon dioxide and other greenhouse gases~\cite{IPCC-AR6}, and that to prevent warming beyond 1.5$^\circ$C, we must limit the emissions of these gases in the future.  In particular, to achieve an 83\% probability that warming will be limited to less than 1.5$^\circ$C, the total carbon budget for the planet is 300 gigatonnes of CO$_2$e\footnote{Carbon dioxide equivalent or CO$_2$e is the number of metric tons of CO$_2$ emissions with the same global warming potential as one metric ton of another greenhouse gas. It is calculated using Equation A-1 in U.S. Code of Federal Regulations Title 40, Section 98.2~\cite{40CFR}. We use the metric ton (t = tonne) throughout this article.}, or about 1.1~t CO$_2$e per year per capita until the year 2050, after which we must reach net zero carbon emissions worldwide.  Current per capita emissions are 4.7~t CO$_2$e worldwide (and also in France and the UK specifically), about 8~t CO$_2$e in Germany, China and Japan, and 14.9~t CO$_2$e in the United States~\cite{OWID-carbon}.  Significant societal action must be taken, starting now, to meet this threat to our current way of life.

Particle physics projects cannot be excluded from these considerations.  While the actions of particle physicists alone cannot stop global warming, the activities of particle physicists have the potential to create greenhouse gas emissions well above that of a typical citizen, and thus well above the per-capita emissions expectations.  
We must anticipate that future accelerators, like any significant public works project, will be scrutinized for their sustainability in a world of reduced carbon emissions.  Physicists will have to be prepared to account for the expected greenhouse gas emissions of their projects, and to take steps to mitigate them.  This will apply to the entire life cycle of an accelerator, from construction through operations through eventual decommissioning.  These mitigations could add extra costs to projects, or cause them to become unfeasible.  While our science seeks to answer basic questions about our world, it must be done in a context that respects the world and protects it for future generations.

In this article, we review the current understanding of the sustainability of future accelerator projects~\footnote{A discussion of emissions associated with current particle physics projects and associated activities, as well as various HEP calculators, can be found in~\cite{HECAP, yHEPcalc, 1point5calc}.}, and highlight steps that can be taken to reduce their associated carbon emissions.  
We devote much of our consideration to the accelerators themselves, since we expect that their construction and operation will lead to the largest amount of CO$_2$e emissions in the life cycle of an accelerator project.  In Section~\ref{sec:colliders}, we will explain the factors that contribute to the CO$_2$e emissions through explicit estimates for a number of proposed colliders.
We also provide briefer summaries of the emissions that might be associated with the operation of potential future particle detectors (Section~\ref{sec:detectors}), the computing facilities required to process and analyze the collected experimental data (Section~\ref{sec:computing}), and travel and other activities necessary to carry out research in large international collaborations (Section~\ref{sec:travel}).  We conclude with suggestions on potential lines of research on mitigating greenhouse gas emissions in particle physics (Section~\ref{sec:RandD}) and closing thoughts (Section~\ref{sec:conclusions}). 

While our focus is on particle accelerators used for high-energy physics research, we note that accelerators are used for many other applications in industry, medicine, chemical and biological research, and defense.  Improvements in sustainability in particle physics might also provide benefits to other fields.

Some terminology common in climate studies is useful in our discussion.  Greenhouse gas emissions by an organization are typically reported in three different {\it scopes}~\cite{scopes}.  Scope 1 covers emissions made directly by an organization; Scope 2 measures emissions from electricity, steam, heat, and cooling acquired from outside the organization, which are sometimes called indirect emissions; and Scope 3 includes all emissions outside of those in the other scopes, such as those associated with daily commuting, work travel, procurement, and catering. 

The {\it global warming potential} (GWP) of a gas refers to how much infrared thermal radiation it would absorb over a given period of time after it has been emitted and is expressed as a factor normalized to the same mass of CO$_2$~\cite{GWP}. 


Studies of the environmental impacts from a project or product use the framework of a {\it life cycle assessment} (LCA).  As the name suggests, an LCA seeks to characterize emissions from the full life of the project, including before use, during use, and end of life.  For civil construction, it is possible to follow engineering standard EN 17472:2022~\cite{LCA-standard}, which labels those three categories as A, B, and C, each with modules for more specific activities.  Category A is the one most commonly considered.

\section{EMISSIONS ASSOCIATED WITH FUTURE ENERGY-FRONTIER COLLIDERS}
\label{sec:colliders}
The most significant piece of infrastructure associated with a future collider program is the collider itself -- the tunnel and the accelerator components that it houses.  The construction and subsequent operations of the collider are thus our strongest area of focus.

We consider the carbon emissions associated with ten possible future energy-frontier colliders, which all address physics goals identified in the most recent P5~\cite{Murayama_2023} and European Strategy~\cite{CERN-ESU-015} reports.  Some of the colliders are explicitly mentioned in those reports, while others have not received as much attention.  We make assumptions on their location based on siting plans described by the individual projects. 
We note that while we are following assumptions of the conditions during construction and operation as currently proposed, these projects are all still evolving.  It is entirely possible that improvements in carbon intensity will be made, especially if the research and development activities described in Section~\ref{sec:RandD} are pursued.  Our evaluation is only a summary of the status at the time of writing.
The projects are:

\begin{itemize}
    \item International Linear Collider (ILC)~\cite{aryshev2023internationallinearcolliderreport}: The ILC is a high-luminosity linear electron positron collider based on superconducting radio-frequency (SRF) technology. The primary components are two linear accelerators (linacs) that point towards a central interaction region.  The initial stage of the ILC would allow for 250~GeV center-of-momentum (CM) collisions, allowing for detailed studies of the Higgs boson.  The necessary site length would be about 20.5~km according to the TDR layout~\cite{ILC-TDR}.  Tohoku, Japan has been the most widely-studied site for the ILC.
    \item Compact Linear Collider (CLIC)~\cite{brunner2022clicproject}: The CLIC design for an $e^+e^-$ collider makes use of a two-beam accelerator concept, in which a high-intensity drive beam, powered by normal-conducting RF cavities, is decelerated to provide energy to the main beam.  The initial stage envisions 380~GeV CM collisions, which would require a tunnel length of 11.4~km.  Siting near CERN in the Geneva, Switzerland area has been considered.
    \item Cool Copper Collider (C$^3$)~\cite{bai2021c3coolroutehiggs}: This $e^+e^-$ linear collider makes use of a distributed coupling accelerator concept that allows for high-gradient operation at liquid nitrogen temperatures.  Collisions at 250~GeV CM would require an 8~km long tunnel.  Unlike other linear collider proposals, the C$^3$ tunnel is designed to be ``cut and cover'', close to the surface rather than deeper underground.  C$^3$ has been proposed by SLAC, thus a United States siting is envisioned.
    \item Future Circular Collider (FCC)~\cite{FCC-overview}: The FCC would be the largest accelerator ever built.  A 90.2~km circumference tunnel would be constructed, sited to connect to the current CERN complex.  The initial phase, FCC-ee~\cite{agapov2022futurecircularleptoncollider}, would provide $e^+e^-$ collisions from 88 to 365~GeV CM energy.  After that 15-year physics program is completed, a hadron collider, FCC-hh~\cite{benedikt2022futurecircularhadroncollider}, would be built in the same tunnel, with the goal of reaching 100 TeV CM proton collisions.  This would yield approximately 10~TeV parton center-of-momentum (pCM) collisions.
    \item Circular Electron Positron Collider (CEPC)~\cite{cepcacceleratorstudygroup2022snowmass2021whitepaperaf3cepc}: This project, based in China, is similar to the FCC, with a 100~km circumference tunnel that would house a $e^+e^-$ collider providing collisions from 91.2 to 360~GeV CM energy.  Also similarly to FCC, a 100~TeV CM proton collider, SppC~\cite{SPPC}, could subsequently be built in the same tunnel.
    \item Muon Collider~\cite{black2023muoncolliderforumreport}: A muon collider is a novel approach to providing 10~TeV pCM collisions.  There would be access to similar physics to that of an $e^+e^-$ collider, but the muon mass is larger than that of the electron, and thus produces less synchrotron radiation, and would require less RF power for acceleration to high energies.  The challenges for such a collider include the short (2.2~$\mu$s) lifetime of the muon, and the backgrounds from muon decay products.  The preliminary design for a muon collider, sited in the United States, includes several tunnels, the largest of which are the accelerator ring, with 17~km circumference, and a collider ring, with a 10~km circumference for 10~TeV CM collisions.
    \item Large Electron Positron Collider 3 (LEP3)~\cite{koratzinos2014lep3lowcosthighluminosityhiggs}: With sufficient improvements to RF cavity power, a new $e^+e^-$ collider could be built in the existing LHC tunnel with access to the $\sim$240~GeV CM energy range. 
    \item High-Energy Large Hadron Collider (HE-LHC)~\cite{HE-LHC}: The development of more powerful magnets would allow for a proton collider of higher CM energy than the current LHC in the existing LHC tunnel. To achieve a near-doubling of the CM energy compared to the LHC, FCC-like 16~T curved magnets would be needed.

\end{itemize}

We separately address emissions arising from the construction and the operation of each collider.  

\subsection{Civil Construction}
\label{subsec:civil}

The greatest amount of carbon emissions is anticipated to arise from construction.  In particular, these contributions are expected to be dominated by those associated with the manufacture of the concrete and steel used in the tunnels. Concrete involves the manufacture of Portland cement~\cite{carbon_brief_cement} which relies on the calcination of limestone in a cement kiln to produce lime.  This chemical process, CaCO$_3$ + heat $\to$ CaO + CO$_2$, naturally releases carbon dioxide into the atmosphere.  Furthermore, the cement kilns are typically heated using fossil fuels.  As a rule of thumb, the production of a tonne of cement is associated with about a tonne of CO$_2$ emissions~\cite{ICE_database}.  Steel production, meanwhile, is highly dependent on coal, which is used as a reducing agent to extract iron from iron ore.  About 1.4 tonnes of CO$_2$ are emitted per tonne of steel manufactured, and it is anticipated that this industry might be among the last to still be using coal in 2050~\cite{steel_IEA}.

\begin{table}
    \caption{\label{Tab:construction}Carbon emissions from civil construction for future colliders.}
    \begin{tabular}{|p{120pt}|p{40pt}|p{140pt}|}

    \hline
        {\bf Collider} & {\bf Emissions (Mt CO$_2$e)} & {\bf Notes (see text for more complete information)} \\ \hline
        ILC (Japan) 250~GeV, 500~GeV & 0.266 & From ARUP report~\cite{ARUP}. \\ \hline
        CEPC (China) 91.2 - 360~GeV & 1.138 & From CEPC presentation~\cite{CEPC-SustainableHEP} which uses the factors of 7.0 kt CO$_2$e/km, 30\% for the auxiliary buildings and 25\% for A4-A5 contributions. \\ \hline
        FCC-ee (CERN) 88 - 365~GeV & 1.056 & From FCC presentation~\cite{FCC-FCCweek}, the deduced emissions per length of the main tunnel is 7.2 kt CO$_2$e/km. \\ \hline
        CLIC (CERN) 380~GeV Drive Beam  & 0.127 & From ARUP report~\cite{ARUP}. \\ \hline
        CCC (USA) 250~GeV, 550~GeV & 0.146 & From CCC paper~\cite{CCC}. \\ \hline
        Muon Collider (USA) 10~TeV & 0.378 & Using 27 km for the sum of the accelerator and collider rings~\cite{black2023muoncolliderforumreport} and using factors of 7.0 kt CO$_2$e/km, 60\% for the auxiliary buildings and 25\% for A4-A5 contributions. \\ \hline
        FCC-hh (CERN) 100~TeV & 0.245 & Re-using the FCC-ee tunnel, using factors of 7.2 kt CO$_2$e/km, 10\% for the auxiliary buildings and 25\% for A4-A5 contributions. \\ \hline
        SPPC (China)  100~TeV & 0.263 & Re-using the CEPC tunnel, using factors of 7.0 kt CO$_2$e/km, 10\% for the auxiliary buildings and 25\% for A4-A5 contributions. \\ \hline \hline
        LEP3 (CERN) 240~GeV & 0.061 & Re-using LHC tunnel, using factors of 6.0 kt CO$_2$e/km, 10\% for the auxiliary buildings and 25\% for A4-A5 contributions. \\ \hline
        HE-LHC (CERN) 27~TeV & 0.061 & Re-using LHC tunnel, using factors of 6.0 kt CO$_2$e/km, 10\% for the auxiliary buildings and 25\% for A4-A5 contributions. \\ \hline
    \end{tabular}
\end{table}

Table~\ref{Tab:construction} gives our tabulation of emissions associated with each of the tunnel and cavern construction for the ten collider projects  considered.  As described below, several of the future collider projects are now collaborating with external firms to provide estimates of emissions, or undertaken their own calculations, using the LCA framework.  Wherever possible, we make use of the projects' own estimates, and for projects that have not completed their own assessments, we have leveraged results from other projects to create similar estimates.  Ultimately, each project will need to complete their own detailed LCA assessments, led by expert engineers in collaboration with project physicists.

The most detailed analysis of expected emissions associated with construction that has been completed so far for potential Higgs factories is the LCA that was completed by ARUP~\cite{ARUP} for CLIC near Geneva and the ILC in Japan.  This LCA considers the impacts associated with the construction of the tunnels, caverns, and access shafts, from raw material extraction through on-site construction activities.  Specifically, the LCA looks at categories A1-A3, which cover emissions associated with materials; and A4-A5, which cover materials transport and construction emissions.  For CLIC, we quote the results for the 380~GeV drive beam option.  For both machines, the dominant source of emissions is in categories A1-A3, which include emissions from the industrial processes used to make concrete and steel needed for the facility.  Note that ARUP analysis yields an estimated emissions per tunnel length of 7.34~kt~CO$_2$e/km for the ILC (which is designed to have a 9.5~m diameter tunnel) and 6.38~kt~CO$_2$e/km for CLIC (which has a 5.6~m diameter tunnel).

C$^3$ has completed an estimate of their emissions from construction, following the ARUP procedure~\cite{PRXEnergy.2.047001}.  Their analysis yields an estimate of the emissions per length of tunnel of 17~kt~CO$_2$e/km, about a factor of two greater than for CLIC, suggesting that further optimizations can be completed.  For calculating the total emissions this is compensated for by the much shorter tunnel length.  

FCC has completed a similar bottom-up estimation of emissions~\cite{FCC-FCCweek}, where we quote their emissions calculations before any material optimization.  It is in $ \sim $10\% agreement with a separate analysis by the C$^3$ team~\cite{PRXEnergy.2.047001} that scales the estimate of the ARUP result of emissions per unit tunnel length for CLIC to the length of the FCC tunnel.  They further estimate that an additional 30\% of emissions over those for the tunnel will arise from auxiliary civil construction such as the caverns and access shafts, and another additional 25\% for A4-A5 type emissions.

CEPC follows the C$^3$ methodology to estimate their emissions from construction, using a factor of 7~kt~CO$_2$e/km for the main tunnel~\cite{CEPC-SustainableHEP}.  This is also within $\sim $10\% agreement of the analysis performed by the C$^3$ team.

For other other possible future colliders, we devise our own emissions estimates.  Neither a HE-LHC nor a LEP3 built in the existing LHC tunnel would create any additional emissions from the main tunnel, as that has already been built and its emissions accounted for.  However, we assume that existing caverns will need to be adapted for detectors that are optimized for the different collisions.  We estimate the carbon impact of the auxiliary civil construction to be 10\% of the LHC tunnel length emissions (assuming 6~kt~CO$_2$e/km as it is similar to the CLIC tunnel) and an additional 25\% for the A4-A5 impacts.

Similarly, FCC-hh (in the FCC-ee tunnel) and SppC (in the CEPC tunnel) would not require the construction of a new main tunnel, but 
additional construction may be required to modify caverns and other areas.  As with the HE-LHC and LEP3, we project an additional 10\% and 25\% of emissions compared to those associated with the main tunnel for the auxiliary civil construction and A4-A5 emissions respectively.

For a muon collider, we take the main tunnels described above, assume  7~kt~CO$_2$e/km emissions for the tunnel, and add an additional 60\% for the auxiliary civil construction and 25\% for the A4-A5 emissions as for the other newly-built facilities.  We use the larger 60\% figure for experimental caverns because while this is a lepton collider, it will produce 10~TeV pCM collisions as FCC-hh and SppC would, and would thus require a similarly large collision hall for the detector.

\subsection{Accelerator Components}

All of the above estimates only consider the global warming potential associated with the construction of tunnels and other auxiliary facilities, and most importantly the production of steel and concrete associated with them.  We also consider the carbon impacts due to the procurement of the accelerator components themselves, the many magnets and RF cavities that are necessary for an energy-frontier collider.  

In a linear collider the dominant structures are the superconducting RF cavities.  As an example and a crude estimate, we consider the cavities needed for the ILC, which are primarily made of niobium. Niobium has very good conductivity, making it suitable for a SRF cavity, but it is  on the U.S. draft list of critical minerals, which ranked niobium second out of 50 critical mineral commodities~\cite{niobium-critical}. A high criticality rating indicates that a ``mineral commodity has important strategic ({\it e.g.}, national security) end uses, but its supply is insecure due to various socioeconomic, geological, and geopolitical factors causing increased supply risk''. Niobium is increasingly used in transport, energy generation and storage and as such demand for niobium is expected to grow in line with countries' targets toward net zero carbon emissions. Based on a description of the cavity geometry~\cite{ILC-TDR} we estimate that each cavity includes a mass of 39.6~kg of niobium. 

The pure niobium needed for these cavities requires significant processing to achieve a low residual resistivity ratio~\cite{aryshev2023internationallinearcolliderreport, XFEL}.  
Information on the carbon intensity of mining and producing pure niobium is limited, although there have been LCA studies on ferroniobium and niobium oxides~\cite{niobiumlca}.  However, the mining and processing of niobium should be similar to that for titanium.  The carbon intensity of titanium alloys is of the order of 50~kg CO$_2$e/kg~\cite{Titanium}.
Although this is a very large carbon intensity compared to other materials mentioned, it is not as large as that of precious metals used in catalytic converters and electronic components, such as palladium (8,500~kg CO$_2$e/kg) and platinum (15,000~kg CO$_2$e/kg).  For our calculations we estimate a carbon intensity of 75~kg CO$_2$e/kg for niobium.

With 8000 cavities expected in the 250~GeV ILC, the total emissions associated with the niobium is about 26~kt CO$_2$e, which is about 10\% of the contribution from the total civil construction.  The cavities therefore have subleading but still significant emissions.  

For a circular collider, such as the FCC-ee, the dominant elements of the accelerator are the magnets.  The 2900 dipole magnets required have 219~kg/m of iron and 19.9~kg/m of aluminum and each dipole has a magnetic length of 24~m~\cite{FCC-ee-CDR}. Assuming a carbon intensity of 2.0~kg~CO$_2$e/kg and 6.8~kg~CO$_2$e/kg for iron and aluminum, respectively~\cite{ICE_database}, the total emissions associated with the materials is about 40~kt~CO$_2$e. Although this is large in absolute terms, it is about 4\% of the emissions associated with the total civil construction, hence also a subleading factor.


\subsection{Operations}
\label{subsec:operations}

Power consumption has long been a concern for accelerator design, as it has historically been the most financially costly element of accelerator operations.  As we aim to operate a sustainable accelerator, we must also consider the cost to the climate due to the carbon emissions associated with power consumption.  In doing so, we  consider the broader context of how power will be generated on the timescale of the operation of future colliders.  All of the projects described in the previous section are at least ten years away from the start of operations.  This may be an underestimate, given that the world's existing collider programs are not yet completed and the limited availability of funding for new construction~\cite{Murayama_2023}.  The Paris Agreement~\cite{paris-agreement} mandates a 43\% decline in greenhouse gas emissions by 2030, far earlier than the operation of any future colliders, which can only be achieved through a substantial de-carbonization of electrical power production.  

An estimation of the carbon intensity of collider operations must start with an assumption on the carbon intensity of the electric power that will be used by each collider.  We follow the ``Announced Pledges Scenario (APS)'' of the International Energy Agency (IEA)~\cite{IEA-energy-outlook, IEA-energy-outlook-2023}, which ``assumes that all national energy and climate targets made by governments are met in full and on time.'' This scenario lies between the other two scenarios considered by the IEA: the ``Net Zero Emissions by 2050 (NZE) Scenario'' and the ``Stated Policies Scenario (STEPS)'' and   we consider it appropriate for our study.  Should nations not meet their announced pledges, global average temperatures are expected to increase by well over 1.5$^\circ$C.  The resulting environmental, social, and political upheavals could be sufficiently great to overwhelm our capacity as a society to build large infrastructures for science, as there will be a more urgent need to address disruptions of climate change.  In short, we ignore less aggressive scenarios for de-carbonization, because if they were to be realized, future colliders might not even be built, let alone operated~\cite{IPCC-AR6-WG2}.  

The expected carbon intensity of electric power varies by region~\footnote{The definition of carbon intensity for electric power depends on whether one considers a scope analysis (emissions from the physical process of producing the electricity) or a LCA analysis (includes the embedded emissions associated with producing the electricity). In this review we use the scope definition,  justified by the fact that in the far future most countries will have negative emissions in place to balance the embedded emissions of electricity producers.}. While all western regions expect to de-carbonize power at about the same rate, the Asian regions (and China especially) currently have higher carbon-intensity power generation than Europe and the United States and thus will take longer to reach zero emissions.  Following the APS scenario from the IEA, we expect that the carbon intensity of electricity production in g CO$_2$e/kWh will proceed as follows:
\begin{itemize}
    \item United States: decrease linearly from 100 to 0 between 2030 and 2040, with fully de-carbonized power afterwards;
    \item Europe: de-carbonized power after 2040;
    \item Japan: decrease linearly from 200 to 50 between 2030 and 2040, and decrease linearly from 50 to 0 between 2040 and 2050;
    \item China: decrease decrease linearly from 200 to 50 between 2040 and 2050, and linearly from 50 to 0 between 2050 and 2060.
\end{itemize}

For each collider considered, we choose the most optimistic date for the start of operations; this would yield the most pessimistic carbon emissions predictions, because it would be earlier in the path towards de-carbonization.  The duration of operations (in years), and the operational power for each phase of operations, is taken from each collider's design document and is summarized in~\cite{PRXEnergy.2.047001}.
We assume that each year includes of $10^7$~s of operating time, a typical choice in projecting accelerator performance~\cite{Roser_2023}.

Table~\ref{tab:operations_emissions} gives the expected carbon intensity of operations for the considered colliders over the physics program for the Higgs boson coupling and total Higgs boson width measurement as described in~\cite{HiggsSnowmass}. We only consider colliders that have any possibility of operating before electrical power is fully de-carbonized worldwide, thus omitting the 10~TeV pCM machines.  As power will be de-carbonized in the United States and Europe by 2040, colliders in those regions will have no indirect emissions due to operations.  For both the ILC and CEPC, the emissions due to operations over the entire collider lifetime are of the same scale as those due to construction (cf. Table~\ref{Tab:construction}). For the case of the CLIC Drive Beam collider, the ARUP LCA analysis provided an estimate of operations emissions (0.06 Mt CO$_2$e)~\cite{ARUP} which is about a factor of two smaller than its construction emissions and much smaller than the Scope 2 operation emissions calculated for the ILC and CEPC.  
In Section \ref{sec:RandD} we discuss the priority from the field to continue the current efforts in the reduction of power and energy from future colliders and detectors.

\begin{table}
    \caption{Carbon emissions due to operations of future colliders using the carbon intensity from the APS scenario of the IEA. The total electrical energy (E$_{\mbox{e}}$) corresponding to the duration of operations for each respective collider is also shown. For the CCC  the numbers in brackets correspond to the optimized power design~\cite{PRXEnergy.2.047001}. A * is inserted to highlight that this value is in the context of a scope analysis~\cite{scopes}.  In the last column it is assumed that electrical power is decarbonized in the U.S. and in Europe, as described in the text.}
    \begin{tabular}{|l|c|c|c|c|c|}
    \hline
        {\bf Collider} & {\bf Start date} & {\bf Duration} & {\bf Total Power} & {\bf Total E$_{\mbox{e}}$} & {\bf Emissions } \\
        & & (y) & (MW) & (TWh) & (Mt CO$_2$e) \\\hline
        ILC (Japan) 250 GeV, 500 GeV & 2035 & 20 & 111, 173 & 7.7 & 0.24 \\ \hline
        CEPC (China) 91.2 - 360 GeV & 2040 & 18 & 283 - 430 & 17.8 & 1.448 \\ \hline
        FCC-ee (CERN) 88 - 365 GeV & 2040 & 14 & 222 - 357 & 11.1 & 0* \\ \hline
        CLIC (CERN) 380 GeV Drive Beam  & 2040 & 8 & 110 & 2.4 & 0* \\ \hline
        CCC (USA) 250 GeV, 550 GeV & 2040 & 20 & 150 (87), 175 (96) & 9.0 (5.1) & 0* \\ \hline
    \end{tabular}
    \label{tab:operations_emissions}
\end{table}

\begin{figure} [htp]
\centering
\includegraphics[width=0.75\textwidth]
{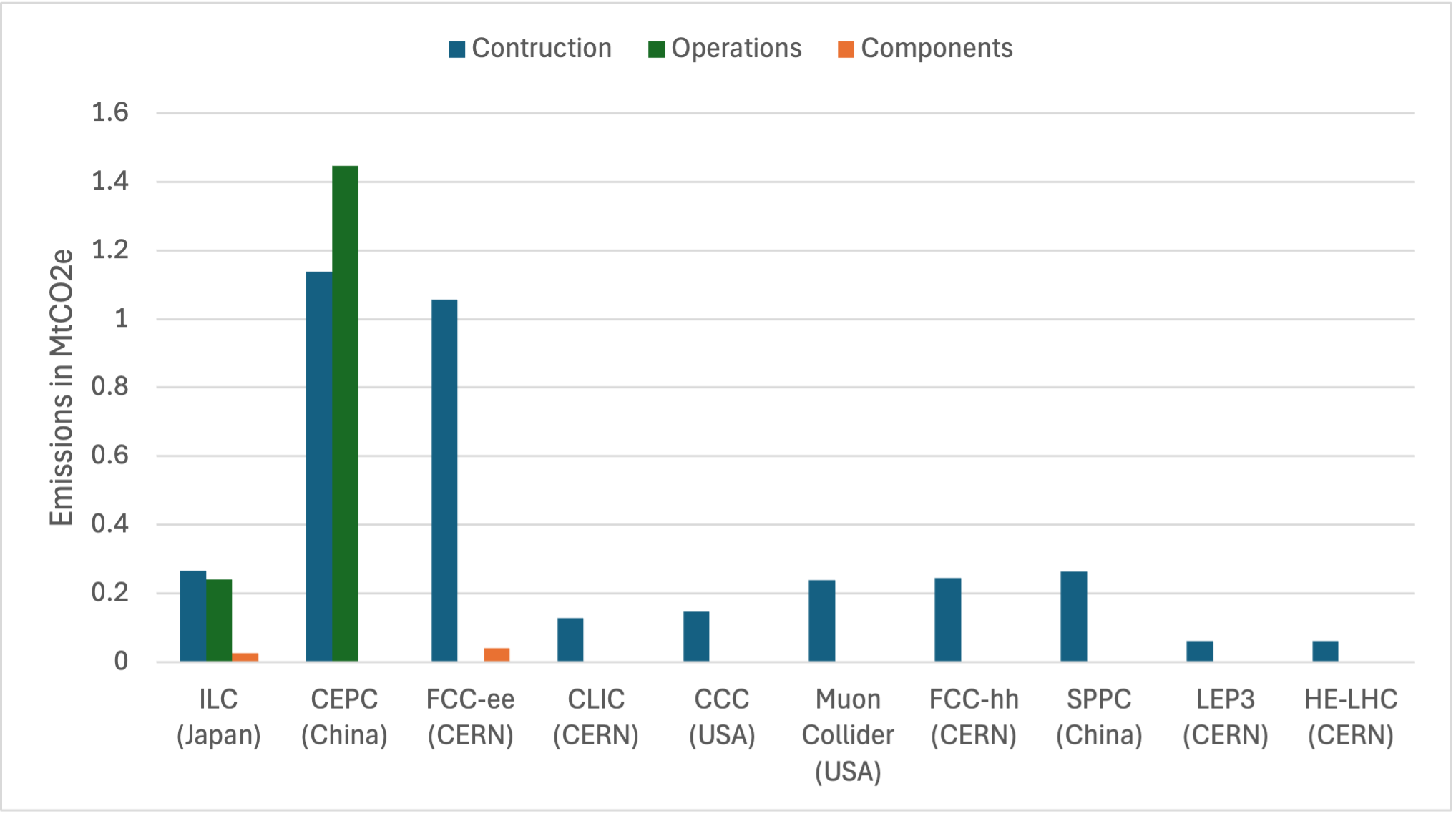}
\caption{\label{Fig1} Emissions of future energy-frontier colliders split according to civil construction, operations and accelerator components. The accelerator components included for representative purposes are the SRF cavitites for the ILC and the dipole magnets from the FCC-ee design. Colliders located in Europe or the U.S. have zero emissions operations from 2040, following the Scope 2 convention defined in~\cite{scopes} and the APS scenario from the IEA. See text for more information on all numbers.}
\end{figure}

In Figure~\ref{Fig1} we show a summary of the emissions of future energy-frontier colliders split according to civil construction, accelerator components and operations. 

\section{EMISSIONS ASSOCIATED WITH POTENTIAL FUTURE DETECTORS}
\label{sec:detectors}
\begin{figure} [htp]
\centering
\includegraphics[width=0.75\textwidth]
{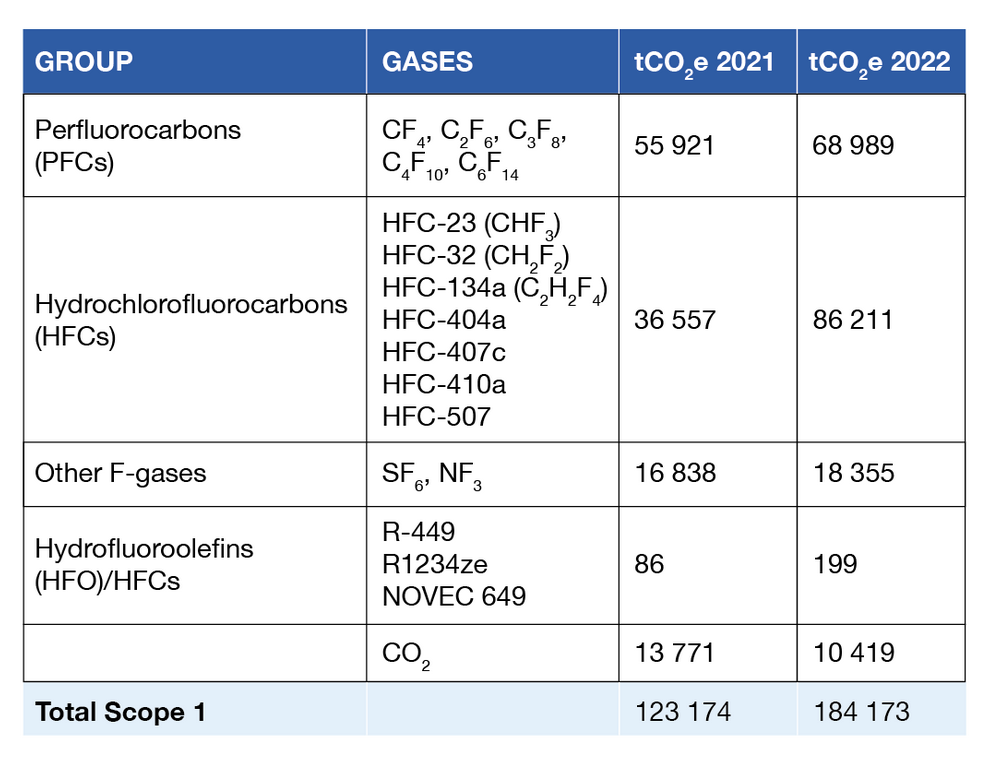}
\caption{\label{Fig-gases} Table of Scope 1 emissions coming from GHG gases used by CERN during 2021-2022~\cite{CERN-environmentreport}.}
\end{figure}

Given that CERN hosts the most powerful accelerator in the world, one might expect that the facility's dominant CO$_2$e emissions arise from electricity generation for the LHC~\cite{CERN-environmentreport}.  Instead, they are from the gases used in the various experiments. For example, as shown in Figure~\ref{Fig-gases}, during 2022, hydrochlorofluorocarbon (HFC) gases contributed 86 thousand tonnes of CO$_2$e emissions released to the atmosphere. The detectors that rely on greenhouse gases (GHG) are used for tracking charged particles. These include Resistive Plate Chambers (RPC), which are typically pairs of parallel plastic plates at an electric potential difference, separated by a gas volume, or Cathode Strip Chambers (CSC), composed of copper strips crossed by arrays of wires in a gas mixture.  Other GHG-dependent detectors are those for particle identification, such as Ring-imaging Cherenkov  (RICH) detectors, where a gas radiator provides information leading to identifying the type of particle that traversed it. Table~\ref{table-gas} lists some of the main GHG used in particle physics detectors, including their 100-year GWPs~\cite{GWP-values}.

\begin{table}[htp]
    \caption{Main GHG used in typical detectors and their associated GWP.}
    \begin{tabular}{|c|c|l|c|}
    \hline
    GHG & Detectors & Properties & GWP \\ \hline
    C$_2$H$_2$F$_4$/R-134a & RPC & primary ionization, charge multiplication & 1300 \\
    SF$_6$ & RPC & electron quenchers & 23500 \\
    CF$_4$ & RICH, MPGD & optical, anti-polymerization, time resolution & 6630 \\
    C$_4$F$_{10}$ & RICH & optical & 9200 \\
    C$_2$F$_6$ & RICH, cooling & radiator & 11100\\
    C$_3$F$_8$ & many & evaporative cooling & 8900 \\
    C$_6$F$_{14}$ & many & liquid coolant & 7910 \\
    \hline
    
        \end{tabular}
    \label{table-gas}
\end{table}

There are two main causes for those gases being emitted into the atmosphere. Some detector gas systems were originally designed and built  to vent those gases directly into the atmosphere when a fresh supply was needed, and some of the detectors are now leaking gases due to cracks in the aged gas inlets. To address the first cause, the CERN gas group has successfully implemented gas recirculation systems over recent years across about 15 detectors operating at CERN~\cite{ICHEP-Rigoletti}. These systems are expensive and complex and require extensive optimization. Nonetheless, in most cases, to maintain optimal detector performance, the gas injected into a detector is about 10\% new gas mixed in with recycled gas, leading to some residual CO$_2$e emissions. An additional approach is also used for some systems, where the gases are recuperated by the gas components being separated at the exhaust and GHG are then stored and re-used as fresh gas. These systems are also quite complex and have to be bespoke for each detector. In general about 60-85\% of the exhausted gases are recuperated with a greater than 90\% quality~\cite{ICHEP-Rigoletti}. 

Energy-frontier detectors that will operate after 2040 will very likely not have access to HFCs or other fluorinated gases, as there is an ongoing phase-down of such gases~\cite{phasedown}. There is already very active R\&D in  particle physics  for finding so-called eco-gas replacements to the GHG used in detectors. For example, at the Gamma Irradiation Facility (GIF) the RPC ECOGas@GIF++ collaboration is performing aging tests using muon beams~\cite{ICHEP-RAMOS}. In general, hydrofluoroolefin (HFO) and perfluoroketone (PKF) are promising molecules to replace GHG. For example, the gas made out of C$_3$H$_2$F$_4$/R-1234ze is a promising replacement to the C$_2$H$_2$F$_4$/R-134a used in RPCs. This gas has a GWP of less than seven. Studies done so far have found that in general a higher high-voltage working point of the detector is needed and higher currents are observed~\cite{EPJPlusRPC}. Unfortunately, some studies have also shown some significant aging issues for detectors using such a gas as part of their mixture~\cite{CBM}.

A potential replacement for the SF$_6$ gas is to use a mixture of Amolea 1224yd and NOVEC 4710~\cite{EPJPlusHallewell} and there are ongoing investigations on this mixture's pollutant potential. A possible replacement for CF$_4$  and C$_4$F$_{10}$  is  either  a combination of C$_5$F$_{12}$ and nitrogen, leading to a reduction of about 45\% in GWP, or to use the NOVEC 5110 gas also in combination with nitrogen. Optical and thermodynamic studies are ongoing using those mixtures~\cite{EPJPlusHallewell}.

Several LHC experiments are moving toward detector cooling systems that rely on CO$_2$. Unfortunately, because of its high triple point, cooling temperatures that can be achieved with a CO$_2$-based system are limited. A successful replacement to the C$_6$F$_{14}$ coolant is the NOVEC 649 product, which has similar thermophysical properties and has good radiation hardness, although it reacts with water which could lead to toxicity issues~\cite{ICHEP-Hallewell}. This product is currently used for cooling the silicon photomultipliers of the scintillating fiber detector of the LHCb collaboration~\cite{LHCbSciFi}.

The NOVEC products from 3M described above are very promising fluoroketone replacements to GHG used in particle physics. Due to their reactivity with atmospheric molecules and their low absorption of infrared radiation, they generally have very low GWP. Unfortunately, these molecules are also part of the per- and polyfluoroalkyl substances (PFAS, so-called ``forever chemicals'') as they are very persistent in soil and water. Such compounds are currently being considered by various governments and could be banned in the near future~\cite{PFAS}.

\section{EMISSIONS ASSOCIATED WITH COMPUTING FOR FUTURE PROJECTS}
\label{sec:computing}
Computing is a key tool of particle physics research, due to the significant data volume generated by experiments, and the high-statistics simulations needed to interpret the recorded data.  Particularly processing-intense workflows include the pattern recognition needed for charged-particle track reconstruction and the simulation of energy loss of particles as they traverse a detector.  In addition, increasingly sophisticated approaches to all elements of computing, including end-user data analysis, are making use of modern software approaches, such as machine learning, that are particularly compute-intensive.

Assessments of computing resource requirements for future collider experiments are still in their infancy, and rightly so given the uncertainties around the evolution of computing technology on the time scale of ten years into the future and beyond.  However, preliminary assessments of computing needs for Higgs factory experiments suggest that the demands will be modest compared to those of the imminent HL-LHC experiments~\cite{Software:2815292}, and in fact on scales already achieved by LHC experiments~\cite{atlascollaboration2024softwarecomputingrun3}.  This is simply due to the fact that $e^+e^-$ collisions produce many fewer particles in the final state compared to collisions of high-intensity proton beams.  In contrast, the computing resource requirements for future hadron colliders at very high energies and instantaneous luminosities would be expected to be significantly greater than those for current experiments, and well beyond even those of the HL-LHC.  Given that such colliders are many decades in the future, and that the very nature of computing is hard to predict on that timescale, we do not consider that case any further.

One estimate of Higgs factory computing needs, for potential ILC experiments, suggests the need for 250~PB of disk storage and 650~kHEPSpec of processing power~\cite{miyamoto_2021_4659567}.  Current rules of thumb~\cite{oli_HCPSS} for the power needs of this much processing and storage lead to an estimate of 930~kW; we can further assume that as computing technology advances, with a continued focus on reducing power consumption, this requirement will be smaller by the time Higgs factories are in operation.

Given the expectation of nearly carbon-free power generation on the timescale of future colliders, we anticipate that the carbon impact of operating the computing resources for these experiments will be negligible in comparison to that of the construction of the collider and detector infrastructure.  Instead, other concerns will come into play.  For instance, as discussed in Section~\ref{sec:RandD}, as human activity becomes increasingly electrified as a means to reduce the use of fossil fuels, the supply of electricity may be increasingly constrained compared to the overall demands.  Therefore, a focus on energy efficiency of computing hardware, and the software algorithms that run on them, will still be needed, even if the energy used is carbon-free.  This can be achieved in part through increasing parallelization of processing, and a transition of processing hardware from traditional central processing units (CPUs) to graphical processing units (GPUs) and tensor processing units (TPUs).  Studies of science application performance at modern supercomputing facilities have demonstrated significant improvements in energy efficiency by using GPUs instead of CPUs for processing, although the impacts vary significantly with the specific application~\cite{Harris_2023}. However, history has repeatedly demonstrated that increased efficiency does not necessarily translates into reduced total energy (Jevons paradox)~\cite{Jevons}, including in the case of the usage of computing farms. 

The dominant concern in the era of de-carbonized power generation will be the global warming impacts associated with the manufacturing process that leads to ``embodied carbon'' in the computing hardware.  This involves the emissions associated with extracting the raw materials (such as heavy metals) that are needed in computers, and the emissions associated with the manufacturing process itself.  For instance, the embodied carbon associated with a typical Dell server used in a data center is about 1300~kgCO$_2$e~\cite{dell_footprint}.  The embodied carbon associated with computer hardware can be reduced through improvements to manufacturing processes, the use of lower-impact materials, and an improved recycling practice of computing components~\cite{10.1371/journal.pcbi.1009324}.  While these are beyond the control of the end users, particle physicists can seek to reduce the impact of embodied carbon by selecting vendors with the most sustainable practices, by trying to extend the lifetime of existing computing hardware, thereby reducing the rate of replacement, and by adhering to strict recycling practices. 

\section{EMISSIONS ASSOCIATED WITH TRAVEL AND OTHERS}
\label{sec:travel}
As a world-wide endeavor, particle physics has historically involved significant air travel by scientists.  While air travel only constitutes about 2.5\% of global CO$_2$e emissions, this share was steadily rising in the period leading up to the COVID-19 pandemic, and will most likely do so again as air travel continues to return to ``normal'' levels~\cite{OWID-airtravel}.  This share will also grow as air travel is expected to be significantly harder to de-carbonize than other sectors of the economy.  While the air travel industry has committed to reach net-zero emissions by 2050, that goal relies on the development of sustainable aviation fuels, which still require significant R\&D for large-scale production and use~\cite{IEA-aviation}.  As 80\% of aviation's emissions are from flights of over 1500~km~\cite{IATA}, there are few alternative modes of transportation that can be used, and electrification of such flights is likely impossible.

Due to the expense of construction and operation, any future energy-frontier collider is likely to be ``one of a kind'' in the world, and given that experimental particle physicists are spread all over the world, many of the participating scientists at the collider will be geographically far away.  This has significant impacts for carbon emissions, simply because emissions largely scale with the distance of the airplane flight.  For instance, the 1604~km round-trip flight from London to Geneva (the site of the proposed FCC) is estimated to produce 418~kg CO$_2$e per passenger, compared to the 4275~kg CO$_2$e per passenger produced for the 14226~km round-trip flight from Chicago to Geneva~\cite{atmosfair}, an order of magnitude difference.

We use this information as inputs to a simple model (see Table~\ref{tab:travel}) of the carbon emissions associated with travel to the experiment site by collaborators, using the FCC as an example.  We assume a collaboration size of 5000, in which half of the collaborators reside near CERN, the experiment site.  One quarter are ``regional'' participants who work London-like distances from CERN and visit the experiment four times per year.  The remaining quarter are ``remote'' participants who live Chicago-like distances from CERN and visit the experiment only two times per year, because of the large distance.  The regional participants produce about 1.7~kg CO$_2$e/year from their air travel, while the remote participants produce about 8.4~kg CO$_2$e/year.  In one year, the collaboration produces 12.6~kt CO$_2$e emissions from air travel, which summed over a 20-year physics program is 0.25~Mt CO$_2$e.  Comparing to the estimates of Section~\ref{sec:colliders}, this amount is more than that associated with the FCC accelerator components, and in fact about one quarter of the emissions associated with the civil construction. This makes travel a subleading but still significant contribution to the carbon emissions over the lifetime of the project.

\begin{table}[htp]
    \caption{A model of experimenter travel to the FCC.}
    \begin{tabular}{|c|c|c|}
    \hline
    & Regional & Remote \\\hline
    CO$_2$e emissions/trip (kg) & 420 & 4275\\
    Collaborators & 1250 & 1250 \\
    Trips/year & 4 & 2 \\
    Emissions/collaborator/year (t) & 1.7 & 8.4 \\\hline
    Total emissions, 20 year program (Mt) & 0.042 & 0.21 \\\hline
   
        \end{tabular}
    \label{tab:travel}
\end{table}

Savings in total emissions can be made by shifting regional air travel to land-based modes such as trains, but much more significant savings can be made by reducing long-haul travel, given the relative difference in emissions per trip.  At the same time, remote collaborators should be able to participate in their experiment as fully as possible, so that they are doing their fair share of work and gaining the experiences and socialization that are important for professional development and scientific progress more generally.  It is thus important to establish regional centers where participants can gather and have the face-to-face interactions that are crucial for collaborative work.  This will require investment in infrastructure such as office space and residential housing from organizations such as national laboratories that would be the natural hosts of these centers.  
Another critical task is the development of the capability for remote operations, so that as much as possible of the regular operation of the experiment can be conducted at a distance.  This includes both the technology needed for the operation of a large-scale detector from far away, and local infrastructure in the form of a remote control room that has a good communication link with the main control room at the experiment site.  Many experiments have already taken such steps, although the motivations have usually been cost savings or local community building rather than environmental concerns.  One successful example is Fermilab LHC Physics Center (LPC)~\cite{LPC}, which hosts many U.S. CMS collaborators for both short- and long-term stays, sponsors workshops on topics related both to CMS and other particle physics areas, and operates a remote control room for the experiment.

It is hard to imagine that long-haul travel can be completely eliminated, especially in the early stages of an experiment where much hands-on work and personal interaction is necessary to establish collaborative and operational  practices.  When travel to the experiment site is needed for operations, it should be for longer trips, {\it e.g.} multi-year stays by staff who will take on long-term operational roles on the experiment.  Given that even a single long-haul trip has a carbon impact much greater than the $\sim 1$ tonne/person yearly quota needed to keep warming to a reasonable level, some form of carbon offset will have to be considered.  

Many routine meetings for experiments are already held by videoconferencing, all the more so since the recent COVID-19 pandemic.  While platforms such as Zoom are functional for many purposes, they do not provide the same quality of interactions as an in-person meeting, and certainly do not provide opportunities to run into a colleague serendipitously in the hallway and exchange ideas.  These features are not unique to particle physics, and this field will most likely take advantages of developments in other industries.

International conferences also have the potential to be a significant source of carbon emissions, as they draw people from all over the planet, many of whom arrive by long-haul airplane flights.  This is issue is not unique to future colliders, and we refer readers to previous discussions of the topic~\cite{bloom2022climateimpactsparticlephysics}.

\section{OBSERVATIONS AND R\&D PRIORITIES}
\label{sec:RandD}

Research and development activities focused on minimizing the environmental impact of future colliders have been in progress for a few years. Several presentations in the 2024 editions of the Sustainable HEP conference~\cite{sustainableHEP} and of the ICHEP conference~\cite{ICHEP} reported on those efforts. Of note was the announcement of the start of the iSAS (innovate for Sustainable Accelerating Systems)~\cite{isas} project funded by EU Horizon 2020. This initiative plans to explore three key innovation directions for energy savings: in the RF system, the cryogenics, and energy savings from the beam itself~\cite{isasSustTalk}. In addition, the CEPC collaboration showed interesting results regarding the power savings which could be achieved using the mid-temperature baking technology, which can enhance the quality factor of the RF cavities by a factor of five compared to the current electropolishing technique~\cite{CEPC-SustainableHEP}. Finally, more projects are using the LCA method to optimize their designs to reduce their environmental impact, as demonstrated by the ISIS-II project~\cite{isis2}. 

The accelerator construction analysis described in Section~\ref{subsec:civil} suggests that the emissions associated with the cement needed for the tunnel could easily dominate the total emissions of a future accelerator, and thus is in the greatest need of mitigation. Consequently, any new accelerator design which minimizes the length of the tunnels has a clear advantage from a carbon emissions and potentially a cost perspective. The recent hybrid, asymmetric, linear Higgs factory design based on plasma-wakefield and RF acceleration is such an example~\cite{HALHF}. As mentioned by the engineering companies involved in the various LCA estimates available, the choice of  material used for the civil construction can result in a wide range of embedded carbon. As an example, in the standard ICE database~\cite{ICE_database} there are 23 different types of cement, which is responsible for the high embedded carbon of concrete. The standard cement assumed by most LCAs is  CEM I, Ordinary Portland Cement, with a carbon intensity of 0.912 kgCO$_2$e/kg. As explained in~\cite{ARUP}, if this cement is mixed in with some Supplementary Cementitious Materials (SCMs), such as fly ash, Ground Granulated Blast furnace Slag (GGBS), limestone powder, or calcined clay, then the carbon intensity may be reduced. For example, the CEM II/B-S - 28\% GGBS which includes between 21-25\% of GGBS material has a carbon intensity of 0.672 kgCO$_2$e/kg, a reduction of 26\% compared to CEM I. However, one has to consider the potential drawbacks of using those alternative materials. These include whether they would be appropriate for a structure  meant to have a lifetime greater than sixty years, and what the cost implications would be, given that the materials are becoming are in increasing demand due to  sustainability concerns worldwide. Finally, there are also very important R\&D avenues being investigated in  sequestering carbon during the cement production stage~\cite{ConcreteColloquium}.

In section~\ref{sec:detectors} the R\&D activities related with the development of new eco-gases to be used in gas-based detectors was discussed. However, given the difficulties mentioned in finding suitable replacements, alternative to gas detectors should also be explored for future energy-frontier colliders. For example, the use of photonic crystals to replace gas radiators in RICH-like detectors is very promising~\cite{crystals}.  For tracking or triggering detectors, one could evaluate the possibility of using plastic scintillators as a replacement for large RPC systems. 

As discussed in section~\ref{sec:computing}, emissions from computing originate from the embedded carbon in the computing components and from the electricity used to power those components. For the latter, as mentioned in section~\ref{subsec:operations}, western countries expect to have a decarbonized electricity grid by 2040, hence a significant fraction of computing farms that will be used to process the energy-frontier collected data will have negligible indirect emissions. However, as already mentioned, demand for electricity is expected to grow significantly due to the necessity for countries to achieve their Net Zero target by 2050. For example, in the U.K., the Climate Change Committee forecasts that electricity demand will double by 2050 compared to 2018 levels~\cite{CCC}, with a similar picture in the U.S.~\cite{EIA}. In addition, installed capacity needs to be at least a factor of two larger than the peak demand for electricity. Achieving this target while only using renewable or nuclear sources will be a significant challenge and thus it is reasonable to assume that the price of electricity will either remain the same as today or more likely increase. As a consequence, it is of vital importance for computing farms and for colliders used for particle physics projects to use as little power and total electrical energy as possible. An interesting area of computing R\&D studies is frequency modulation, where the CPU clock can run at different rates and be dynamically adjusted to adapt to the power demand of the grid~\cite{ATLASICHEP}. 

Some authors have suggested that one possible avenue to ensure that renewable electricity sources are used for the power needs of future facilities is for those facilities to purchase such renewable providers ({\it e.g.} solar farms or wind turbine farms). It has also been suggested that CERN could get its electricity from solar farms in North Africa, transmitted via sub-sea cables (see Case Study 3.3 in~\cite{HECAP}). First it is important to note that Scope 2 (indirect) emissions do not include the emissions associated with the construction and the embodied carbon of the electricity sources, while if a facility builds or purchases such a source, those emissions would need to be included in their Scope 1 emissions. Finally, we believe that the implications of particle physics research facilities acquiring such resources should be explored and discussed. Given the climate emergency facing humanity, particle physics researchers should instead contribute to the work needed for every country to decarbonize their electricity grid, supporting the minimization of Scope 2 emissions for particle physics projects.  

As discussed in section~\ref{sec:travel}, one way to mitigate travel emissions is to maximize the use of videoconferencing systems. However, as experienced during the COVID-19 pandemic, this does not capture the spontaneous interactions crucial to the exchange of ideas, and it can also lead to online fatigue. The field of particle physics has a long tradition of being an innovator in collaborative tools, such as the invention of the World Wide Web. This spirit of innovation should continue to enable minimum travel. This will be especially crucial given that decarbonizing flight is a very difficult goal to achieve and, some countries have Green Parties advocating legislation aimed at introducing a frequent-flyer levy~\cite{UKGreen}. The field of particle physics could  champion a vast expansion of the use of Virtual/Augmented Reality headsets.  These can help make online social interactions more natural and also support participation from the Global South in particle physics projects.

\subsection{Negative emissions technologies}

Given the social context associated with the climate emergency, especially for projects being built in the 2030's and operating during the 2040's, it is reasonable to predict that such large projects will be required to satisfy a net zero policy.  Some particle physics-related facilities already have such a program~\cite{STFZ-NZ}. Despite all of the attempts described above to reduce emissions from future energy-frontier facilities, it is very likely that such projects will still emit a significant amount of emissions. Consequently, negative emissions technologies might need to be considered and actively pursued by particle physics researchers, thereby benefiting society in the process. 

Negative emissions technologies can be separated into two broad categories: nature-based solutions and carbon dioxide capture~\cite{CarbonBrief-Negative}. Note that Carbon Capture and Storage (CCS) is not a negative emissions technology, as it aims to capture the CO$_2$ directly at the source, so its contribution is carbon neutral rather than carbon negative~\cite{CCS}.  Standard nature-based technologies for negative emissions, as discussed by the Green ILC~\cite{GreenILC} project, is afforestation (planting new trees) or re-forestation  (reintroducing a forest), which absorbs CO$_2$ as part of the photosynthesis process. Note that we do not consider protecting an already existing forest from being cut down in the future to be net negative in terms of emissions. Trees absorb CO$_2$ according to their species and age. A juvenile tree (15 years old) will start to reach peak absorption until it reaches the mature stage (50 years old) where, depending on the species, it could absorb between 200-400 kg of CO$_2$ per year~\cite{trees}. In comparison, during the sapling stage (5-15 years old), the tree would absorb between 10-50 kg of CO$_2$ per year. Another important aspect to consider in using afforestation is that the overall carbon cycle of a tree during its whole life span is carbon neutral; all of the carbon absorbed during its life will be released into the atmosphere when the tree dies and decays. For trees to be a negative emissions technology, they need to be cut and used, for example in construction, so as to sequester the carbon. If the trees are burned and used as a source of fuel, then CCS needs to be utilized to capture those emissions. In order to absorb 1 million tonnes of CO$_2$ emissions over a period of 10 years (an amount equivalent to that which would be released through the construction of the FCC-ee tunnel), we estimate that about 10 million new trees would be needed. These would cover an area between 4,000 to 10,000 hectares of land, depending on various factors including the tree species. This area is equivalent to a factor 10 to 30 of the area covered by Central Park in New York City, or four to 11 times the area of Richmond Park in the London area. 


The second approach to negative emissions technologies is that of carbon capture, in particular Direct Air Capture (DAC). There are currently two main methods for DAC.  One uses a hydroxide solution to capture CO$_2$ from ambient air, which is then heated to high temperatures (using renewable power) to release the CO$_2$ so it can be stored and the hydroxide solution reused.  The other method uses amine adsorbents in small, modular reactors; the temperature needed to re-release the CO$_2$ is estimated to be less than for the first method~\cite{CarbonBrief-DAC}. The latter method is used by the Swiss firm Climeworks~\cite{Climeworks}. It currently operates the Orca DAC plant in Iceland which can capture 4,000 tonnes of CO$_2$ per year.  A larger facility, Mammoth, which will capture four times this amount per year, is currently being built with construction costs estimated at \$600M. Climeworks uses between 8,000 to 18,000 kWh of electrical energy to capture 1 tonne of CO$_2$. In Iceland this power is provided by the Hellishei\dh i geothermal facility, which can produce 2.65 TWh of electricity per year. It would then take Mammoth 28 years of operation in order to absorb 1 million tonnes of CO$_2$ emissions from the atmosphere.


\section{SUMMARY AND CONCLUSION}
\label{sec:conclusions}


In this article, we reviewed the emissions of future energy-frontier colliders, putting particular emphasis on their civil construction and operations. We reported that some major accelerator projects have already started collaborations with engineering companies to develop state-of-the-art LCA analyses. Taking inspiration from those analyses, we estimated civil construction emissions for other potential energy-frontier colliders operating in the 2040's. Not surprisingly, the larger footprints of colliders are associated with higher emissions. On the operations side, China and Japan have a longer timeline for de-carbonizing their electricity grids compared to Europe and the U.S. and we estimated the operations emissions of the experiments planned to be built in those countries. In particular for CEPC in China, those operations emissions are of the same order of magnitude as the civil construction associated with building a 100 km tunnel. We reiterated the crucial importance for accelerators to minimize their operating power. 

It is likely that gas detectors will continue to be an attractive option as part of the future detectors associated with those energy-frontier colliders. Unfortunately, the search for gases with low GWP and acceptable toxicity is challenging and it is imperative that the particle physics detector community  fully engage in this R\&D. Suitable replacement of those detectors with non-gas options should also be explored. The computing section highlighted the growing importance of the embodied carbon in computer components as power continues to become carbon free, while the travel section conveyed the importance of reducing travel and finding innovative ways to pursue collaborative research.

The central message of these findings is that significantly reducing emissions or adhering to a net zero target for these future projects (under current research budgets) will be extremely challenging. The physics program of all these projects demonstrate how vibrant, creative and exciting the future of the field is.  In order to secure this bright future it is  imperative for the particle physics community to take actions for sustainability that go beyond our traditional accelerator and detector activities. Although the techniques associated with, for example, producing low carbon cement and direct air capture have been demonstrated, their application at scale is far from guaranteed and requires technological innovation that the particle physics community can positively contribute to. Our future depends on it.

\section*{DISCLOSURE STATEMENT}
The authors are not aware of any affiliations, memberships, funding, or financial holdings that
might be perceived as affecting the objectivity of this review. 

\section*{ACKNOWLEDGMENTS}
KB is supported by National Science Foundation award 2209764, while VB is supported by the UK Science and Technology Facilities Council award ST/W000555/1. We would like to thank the following colleagues for useful discussions: Daniel Britzger, Yann Coadou, Oliver Gutsche, Valerie Lang, Kristin Lohwasser,  Peter Millington, and David Waters.

%

\end{document}